# SPEECH ENHANCEMENT USING KERNEL AND NORMALIZED KERNEL AFFINE PROJECTION ALGORITHM

Bolimera Ravi [1] and T. Kishore Kumar [2]

[1] Department of ECE, Vignan University, Vadlamudi, Andhra Pradesh, 522213.
`ravi_gamr@yahoo.com`
[2] Department of ECE, NIT Warangal, Warangal, Andhra Pradesh, 506004.
`kishoret@nitw.ac.in`

## ABSTRACT

*The goal of this paper is to investigate the speech signal enhancement using Kernel Affine Projection Algorithm (KAPA ) and Normalized KAPA. The removal of background noise is very important in many applications like speech recognition, telephone conversations, hearing aids, forensic, etc. Kernel adaptive filters shown good performance for removal of noise. If the evaluation of background noise is more slowly than the speech, i.e., noise signal is more stationary than the speech, we can easily estimate the noise during the pauses in speech. Otherwise it is more difficult to estimate the noise which results in degradation of speech. In order to improve the quality and intelligibility of speech, unlike time and frequency domains, we can process the signal in new domain like Reproducing Kernel Hilbert Space (RKHS) for high dimensional to yield more powerful nonlinear extensions. For experiments, we have used the database of noisy speech corpus (NOIZEUS). From the results, we observed the removal noise in RKHS has great performance in signal to noise ratio values in comparison with conventional adaptive filters.*

## 1. INTRODUCTION

Adaptive filtering is an important subfield of digital signal processing having been actively researched for more than five decades and having important applications such as noise cancellation, system identification, etc.. In such noise removal applicable systems, the signal characteristics are quite faster rate. Recent research on adaptive filter has focused on a high dimensional to yield powerful non-linear approach of the signals. Kernel adaptive filters made it easier through Reproducing Kernel Hilbert Space (RKHS). This approach has been justified due to the variation of Signal-to-Noise Ratio (SNR) across the speech spectrum.

Unlike the White Gaussian Noise (WGN) has a flat spectrum, but the real world noise spectrum is not flat. Thus, the effect of noise signal over speech signal is non-uniform over the whole spectrum. Some are affected more adversely than the other frequencies. In multi-talker babble, most of the speech energy resides at the low frequencies, are affected more than the high frequencies. Hence to estimate a suitable factor that will subtract just the necessary amount of the noise spectrum from each frequency bin (ideally) becomes imperative, to prevent corrosive subtraction of the speech while residual noise reduction. One more factor that leads to variation in SNR is the fact that noise has random effect on different vowels and consonants.

Due to the random nature of the noise and the inherent complexities of speech signal, it is more difficult to get rid of various types of noise. Noise reduction techniques usually have a tradeoff between the amount of noise removal and speech distortions. Many noise reduction algorithms have been proposed to reduce the noise effect for speech enhancement, for example, Spectral Subtraction (SS), Wiener Filtering, Minimum Mean Square Error (MMSE) based speech estimation, Kalman Filtering and Bayesian based estimation etc.

## 2. SPEECH ENHANCEMENT

Speech Enhancement is a challenging task in many applications like hearing aids, forensic applications, military, cellular environments, front-ends for speech recognition system, telecommunication signal enhancement, etc. It is simply the improvement in intelligibility and/or quality of a degraded speech signal by using signal processing tools. By speech enhancement, it comprises not only to noise reduction but also to dereverberation and separation of independent signals.

Speech Enhancement is a fundamental for research in the applications of digital signal processing. Speech Enhancement is a very difficult problem mainly for two reasons. Firstly, the characteristics and nature of the noise signals can change dramatically in time and for various applications and also to find corresponding algorithms that really work in various practical environments. Secondly, for each application, the performance measure can also be defined differently. The following Figure 1 shows the basic idea of speech enhancement.

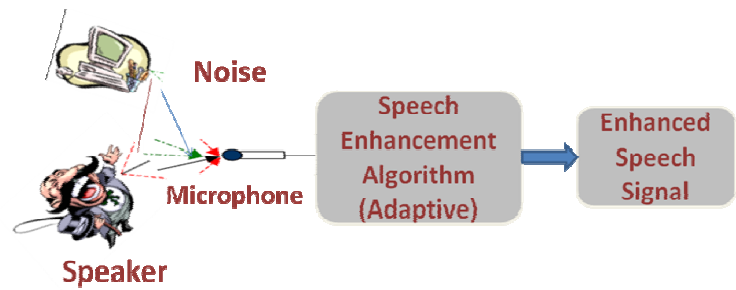

Figure 1. Basic idea of Speech Enhancement

## 3. ADAPTIVE FILTERS

Whenever there are either unknown fixed specifications or unsatisfied specifications by time-invariant filters, an adaptive filter is required. Since the characteristics are dependent on the input signal, an adaptive filter is a nonlinear filter and consequently the homogeneity and additivity conditions are not satisfied. Adaptive filters are time-varying since filter parameters are continually changing to meet performance requirement. Adaptive filters have been effective and popular approaches for the speech enhancement for the past decade.

Before starting the APA and KAPA, an adaptive filter introduction is given as follows. The name "adaptive filters", adjective adaptive can be empathized by considering the system which tries to adjust itself so as to respond to some surrounding phenomenon. In other words the system tries to adjust its parameters with an aim of meeting some well define target which depends on system state as well as its surrounding. Furthermore there is a need to have certain procedure by which the process of adaptation is carried out. And finally the system which undergoes the process of adaptation is called by the more technical name "filter".

The adaptive filter has advantages like lower processing delay and better tracking of the trajectory of non-stationary signals [2]. These are essential characteristics in applications such as noise estimation, echo cancellation, adaptive delay estimation and channel equalization, where low delay and fast racking of time-varying environments and time-varying processes are important objectives.

The existence of a reference signal which is hidden in the fixed-filter approximation step, defines the performance criterion. The general adaptive-filter configuration is shown in figure 2.

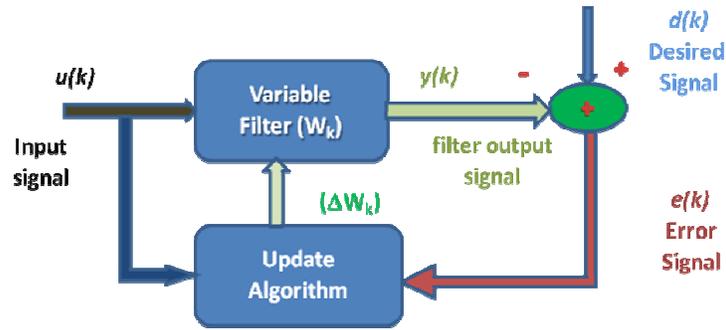

Figure 2. General Adaptive filter configuration.

An adaptive filter is specified as

$$y(k) = w^T(k)\, u(k) \qquad (1)$$

Where k is the time index, y is the filter output, x the filter input, Wk are the filter coefficients. The block diagram, shown in the following figure 2, serves as a foundation for particular adaptive filter realization [11], such as Affine Projection Algorithm (APA). The theme behind the general configuration is that a variable filter that extracts an estimate of the desired signal.

The following assumptions were made:

Input signal u(k) = desired signal d(k) + interfering noise v(k)

$$u(k) = d(k) + v(k) \qquad (2)$$

The variable filter is having a Finite Impulse Response (FIR) structure. i.e. for such structures the impulse response is equal to the filter coefficients. For a filter of order p, the coefficients are defined as

$$W_k = [W_k(0),\, W_k(1),\, \ldots\ldots W_k(p)] \qquad (3)$$

The difference between the desired and the estimated signal is the error signal or cost function

$$e(k) = d(k) - y(k) \qquad (4)$$

The variable filter will estimates the desired signal by convolving the input signal along with the impulse response. The expression in vector notation as

$$y(k) = W_k * u(k) \qquad (5)$$

Where

Signal & Image Processing : An International Journal (SIPIJ) Vol.4, No.4, August 2013

$$u(k) = [u(k), u(k-1),...u(k-p)] \qquad (6)$$

is an input signal vector. Furthermore, the variable filter updates its coefficients at every instant of time

$$W_{k+1} = W_k + \Delta W_k \qquad (7)$$

Where $\Delta W_k$ is a correction factor for the filter coefficients. Based on the input and error signals, the adaptive algorithm generates this correction factor

### 3.1. Affine Projection Algorithm (APA)

Affine Projection algorithm (APA) [10] was derived as a generalization of the NLMS algorithm. In APA, the projections were made in multiple dimensions where as one dimensional in NLMS, in the sense that each tap weight vector update of the NLMS is viewed as a one dimensional affine projection, while in the APA multiple dimensional projections were made. As increasing the projection dimension, increases the convergence rate of the tap weight vector. However, it leads to an increased computational complexity.

Let d (desired signal) be a zero-mean scalar-valued random variable and let u (noisy signal) be a zero-mean L × 1 random variable with a positive-definite covariance matrix $R_u = E[uu^T]$. The cross-covariance vector of d and u is denoted by $r_{du} = E[du]$. The weight vector w that solves

$$\text{Min}_w (E |d - w^T u|2 ) \qquad (8)$$

is given by $w^0 = R_u^{-1} r_{du}$ [3].

Several methods that approximate 'w' iteratively also exist. For example, the common gradient method

w(0) = initial guess;

$$w(k) = w(k-1) + \eta[r_{du} - R_u w(k-1)] \qquad (9)$$

or the regularized Newton's recursion,

w(0) = initial guess;

$$w(k) = w(k-1) + \eta(R_u + \varepsilon I)^{-1} [r_{du} - R_u w(k-1)] \qquad (10)$$

where $\varepsilon$ is a small positive regularization factor and $\eta$ is the step size specified by designers. Stochastic-gradient algorithms replace the covariance matrix and the cross-covariance vector by local approximations directly from data at each iteration. To obtain such approximations, several ways are available. The trade-off is convergence performance, computation complexity and steady-state behavior [3].

Assuming that we have access to random variables (d and u) observations over time

{d(1), d(2), . . . } and {u(1), u(2), . . . }

The Least-mean-square (LMS) algorithm simply uses the instantaneous values for approximations $\hat{R}_u = u(k)u(k)^T$ and $\hat{r}_{du} = d(k)u(k)$. The corresponding steepest-descent recursion (9) and Newton's recursion (10) become

$$w(k) = w(k-1) + \eta u(k)[d(k) - u(k)^T w(k-1)] \qquad (11)$$

$$w(k) = w(k-1) + \eta u(k)[u(k)^T u(k) + \varepsilon I]^{-1}[d(k) - u(k)^T w(k-1)] \qquad (12)$$

The affine projection algorithm however employs better approximations. Specifically, the approximations $R_u$ and $r_{du}$ are replaced by the instantaneous approximations from the K most recent regressors and observations. Denoting

$U(k) = [u(k - K + 1), ..., u(k)]_{L \times K}$ and

$d(k) = [d(k - K + 1), . . . , d(k)]^T$

one has

$$\hat{R}_u = (1/K)U(k)U(k)^T \text{ and } \hat{r}_{du} = (1/K)U(k)d(k) \qquad (13)$$

Therefore (9) and (10) become

$$w(k) = w(k-1) + \eta U(k)[d(k) - U(k)^T w(k-1)] \qquad (14)$$

### 3.2. Normalized Affine Projection Algorithm (NAPA)

The normalized affine projection algorithm becomes

$$w(k) = w(k-1) + \eta [U(k)^T U(k) + \varepsilon I]^{-1} U(k)[d(k) - U(k)^T w(k-1)] \qquad (15)$$

and (15), by the matrix inversion lemma, is equivalent to [3]

$$w(k) = w(k-1) + \eta U(k)[U(k)^T U(k) + \varepsilon I]^{-1}[d(k) - U(k)^T w(k-1)] \qquad (16)$$

It is noted that this equivalence lets us deal with the matrix $[U(k)^T U(k) + \varepsilon I]$ instead of $[U(k)U(k)^T + \varepsilon I]$ and it plays a very important role in the derivation of kernel extensions. We call recursion (14) APA and recursion (16) normalized APA.

## 4. KERNEL ADAPTIVE FILTERS

Kernel method is a good nonparametric modeling technique. The main theme of kernel method is to transforming the input data into a high dimensional feature space through a reproducing kernel such that the inner product operation can be computed efficiently in the feature space through the kernel evaluations [5]. After that an appropriate linear methods are subsequently applied on the transformed data. As long as an algorithm can be formulated in terms of equivalent kernel evaluation (or inner products), there is no call to perform computations in the high dimensional feature space. This is the main advantage when compared to the traditional methods. Successful examples of this methodology include kernel principal component analysis, support vector machines (SVM's), etc. The SVM's have already shown good functioning in increasing the accuracy of speech recognition activity.

Kernel adaptive filters are online kernel methods, closely related to some artificial neural networks such as radial basis function networks and regularization networks[6]. The kernel adaptive filtering technique used in this work for general nonlinear problems. It is a natural generalization of linear adaptive filtering in reproducing kernel Hilbert spaces.

### 4.1. Kernel Affine Projection Algorithm

A kernel [11] is a symmetric, continuous, positive-definite function $\kappa : U \times U \rightarrow R$. U is the input domain, a compact subset of RL. The commonly used kernels include the Gaussian kernel (17) and the polynomial kernel (18):

$$\kappa(u, u') = \exp(-a\|u - u'\|^2) \quad (17)$$

$$\kappa(u, u') = (u^T u' + 1)^P \quad (18)$$

Steps involved in kernel affine projection algorithm are as follows

---

**Algorithm : Kernel Affine Projection Algorithm (KAPA)**

---

**Initialization:**
learning step η

$$a_1(1) = \eta d(1) \quad (19)$$

while {u(k), d(k)} available do

    % allocate a new unit of weight vector

$$a_k(k-1) = 0 \quad (20)$$

for n = max(1, k − K + 1) to k do

    % evaluate outputs of the current network

$$y(k,n) = \sum_{j=1}^{k-1} a_j(k-1)\kappa_{n,j} \quad (21)$$

$\kappa$ is the kernel function.

    % computer errors

$$e(k, n) = d(k) - y(k, n) \quad (22)$$

    % update the min(k,K) most recent units

$$a_n(k) = a_n(k-1) + \eta e(k, n) \quad (23)$$

end for

if k > K then

    % keep the remaining

    for n = 1 to k − K do

$$a_n(k) = a_n(k-1) \quad (24)$$

    end for

end if

### 4.2. Normalized Kernel Affine Projection Algorithm

Normalization factor of APA is $G(k) = [U(k)^T U(k) + \varepsilon I]^{-1}$

For kernel APA it becomes

$$a_n(k) = a_n(k-1) + \eta e(k,n)g(k) \quad (25)$$

## 5. EXPERIMENTAL RESULTS

In this section, we investigate the performance of the KAPA and APA algorithms for speech enhancement at various signal-to-noise ratios of different noise conditions. The step size used for both the APA and KAPA is 0.2. The separate noise corpus from NOIZEUS [9] were collected and added to the clean Speech signals for the experimentation. At different noisy levels, performances of these evaluated for speech enhancement. Babble noise, Train noise, White noise, Car noise and Restaurant noise at 0, 5, 10, and 15 dB SNR were experimented. For this work, a total of 20 datasets were generated.

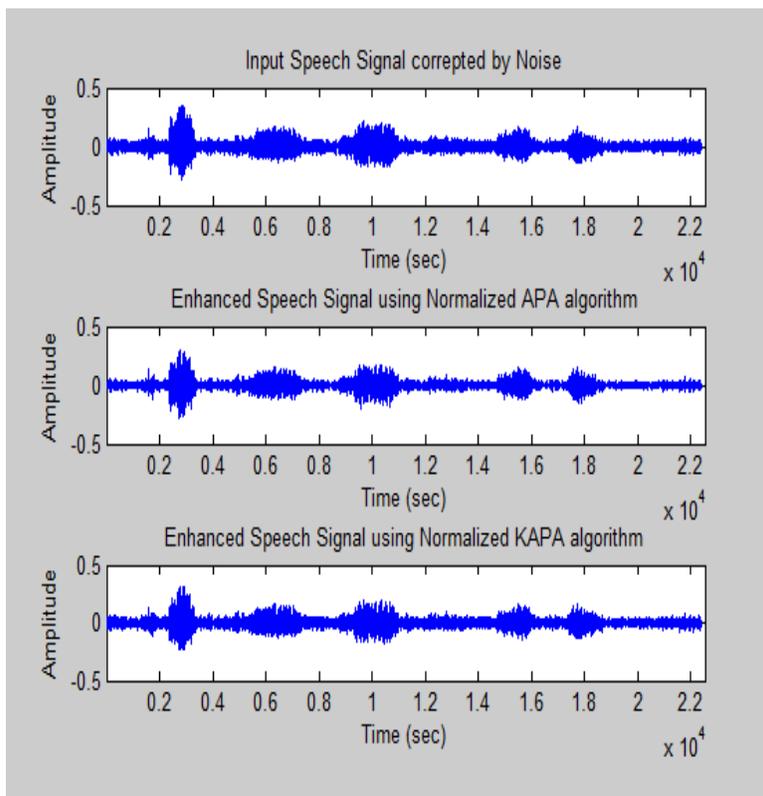

Figure 3. Speech signal of, a) Speech corrupted by Car Noise of 5dB, b) Recovered speech signal using APA algorithm, c) Recovered Speech Signal using Normalized KAPA algorithm.

The above figure 3 shows the original speech signal, APA and KAPA Enhanced speech signals which are corrupted by car noise of 5dB SNR The below figure 4 shows the Spectrograms of original speech signal, APA and KAPA Enhanced speech signals which are corrupted by car noise of 5dB SNR.

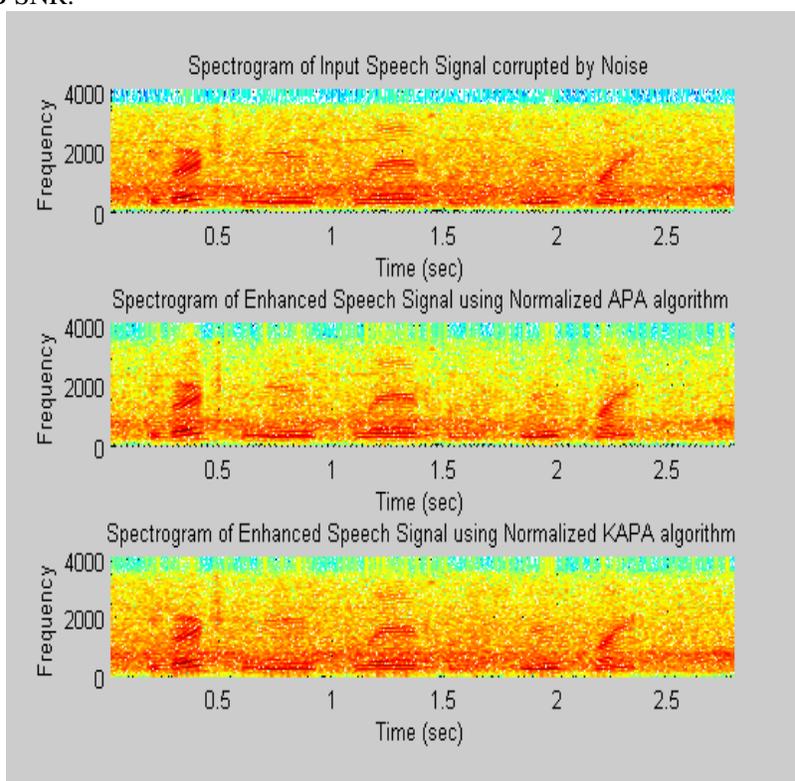

Figure 4. Spectrograms of, a) Speech corrupted by Car Noise of 5dB, b) Recovered speech signal using Normalized APA algorithm, c) Recovered Speech Signal using Normalized KAPA algorithm

## 6. PERFORMANCE ANALYSIS

The main objective of the adaptive filters is the error signal e(k) minimization. Its success will clearly depends on the length of the adaptive filter, the nature of the input signals, and the adaptive algorithm used. The signal is perceived by listeners reflects the subjective measure of quality of speech signals. At 0 dB the two signals are of equal strength and positive values are usually connected with better intelligibility where as negative values are connected with loss of intelligibility due to masking. Positive and higher SNR values are found in all the algorithms. The performances are measured based on the metrics namely MSE and SNR for all the algorithms.

### 6.1. Mean Square Error (MSE)

MSE is defined as 'mean of error squares' and is calculated using the formula

$$MSE = \frac{\sum (y_i - \hat{y}_i)}{n - p} \tag{26}$$

In order to quantify the difference between values implied and the true being estimated, the MSE of an estimator is used.

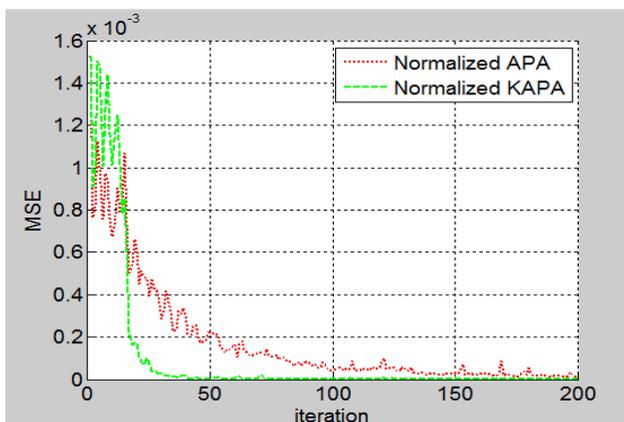

Figure 5. MSE (Learning Curves) comparison of Normalized 'APA and KAPA' algorithms for a Speech signal corrupted by Car Noise of 5dB

## 6.2. Signal to Noise Ratio (SNR)

Signal to Noise Ratio (SNR) is defined as the ratio of power between the signal and the unwanted noise. SNR is calculated using the formula

$$\frac{S}{N} = \frac{n_{signal}}{n_{ratio}} \qquad (27)$$

One of the most important goals of any speech enhancement technique is to achieve highest possible SNR. Higher the SNR ratios, better the performance of speech signal enhancement. In given table 1, we have shown the comparison of SNR values for the experimented algorithms APA and KAPA respectively

Table 1. Comparison of SNR values for APA and KAPA algorithms

| Noise Type | SNR (dB) | Enhancement Method | |
|---|---|---|---|
| | | APA | KAPA |
| Babble Noise | 0 | 11.5697 | 23.6469 |
| | 5 | 6.2768 | 20.6057 |
| | 10 | 4.7025 | 18.1252 |
| | 15 | -0.3462 | 12.4301 |
| Train Noise | 0 | 13.2165 | 23.1598 |
| | 5 | 12.7766 | 25.6758 |
| | 10 | 1.1339 | 15.1536 |
| | 15 | -2.5412 | 11.5884 |
| Car Noise | 0 | 14.5846 | 24.7101 |
| | 5 | 9.2950 | 22.8052 |
| | 10 | 2.3719 | 16.3343 |
| | 15 | -0.7025 | 12.4383 |
| Restaurant Noise | 0 | 10.2245 | 21.1858 |
| | 5 | 9.9622 | 23.9125 |
| | 10 | 0.0635 | 13.2941 |
| | 15 | 3.8029 | 16.2593 |

Signal & Image Processing : An International Journal (SIPIJ) Vol.4, No.4, August 2013

## 7. CONCLUSIONS

The speech communication system performs greatly when the input signal has no limits or no noise effects and is degraded when there is a fairly large level of noise input signal. In such cases system cannot meet speech intelligibility, speech quality, or recognition rate requirements. In this paper, a new methodology has been applied for speech enhancement with RKHS. We observed that there is a better improvement in removal of background noise in RKHS. The KAPA algorithm has also shown much better noise ratio values compared to the APA algorithm.

Signal & Image Processing : An International Journal (SIPIJ) Vol.4, No.4, August 2013

## 7. CONCLUSIONS

The speech communication system performs greatly when the input signal has no limits or no noise effects and is degraded when there is a fairly large level of noise input signal. In such cases system cannot meet speech intelligibility, speech quality, or recognition rate requirements. In this paper, a new methodology has been applied for speech enhancement with RKHS. We observed that there is a better improvement in removal of background noise in RKHS. The KAPA algorithm has also shown much better noise ratio values compared to the APA algorithm.

## REFERENCES

[1] W. Harrison, J. Lim, E. Singer, "A new application of adaptive noise cancellation," IEEE Trans. Acoustic Speech Signal Processing, vol.34, pp. 21-27, Jan 1986.
[2] SaeedV.Vaseghi," Advanced digital signal processing and noise reduction." John Wiley and Sons.
[3] A. Sayed, Fundamentals of Adaptive Filtering. New York: Wiley, 2003.
[4] http://en.wikipedia.org/wiki/Aadaptive_filter
[5] W. Liu, P. Pokharel, and J. C. Pr´ıncipe, "The kernel least mean square algorithm," IEEE Transactions on Signal Processing, vol. 56, 2008.
[6] http://en.wikipedia.org/wiki/Kernel_adaptive_filter
[7] Jose C. Principe, Weifeng Liu, Simon Haykin "Kernel Adaptive Filtering: A Comprehensive Introduction", Wiley, March 2010.
[8] Weifeng Liu and J. Principe, "Kernel Affine Projection Algorithms," EURASIP Journal on Advances in Signal Processing, vol. 2008, Article ID 784292, 12 pages, 2008. doi:10.1155/2008/784292
[9] http://www.utdallas.edu/~loizou/speech/noizeus/index.html
[10] Constantin Paleologu, Jacob Benesty, Silviu Ciochinˇa, "Sparse Adaptive Filters for Echo Cancellation", 2010, Morgan and Claypool publishers.
[11] N. Aronszajn, "Theory of reproducing kernels," Trans. Amer. Math. Soc., vol. 68, pp. 337–404, 1950.